\documentclass[conference]{IEEEtran}
\IEEEoverridecommandlockouts
\usepackage{cite}

\usepackage{adjustbox}
\usepackage{amsmath,amssymb,amsfonts}
\usepackage{graphicx}
\usepackage{textcomp}
\usepackage{xcolor}
\usepackage{multicol}
\usepackage{multirow}
\usepackage{subcaption}
\usepackage{enumitem}
\usepackage{balance}
\usepackage{booktabs}
\usepackage{array}
\newcolumntype{x}[1]{>{\centering\arraybackslash\hspace{0pt}}p{#1}}

\newcommand\blfootnote[1]{%
  \begingroup
  \renewcommand\thefootnote{}\footnote{#1}%
  \addtocounter{footnote}{-1}%
  \endgroup
}

\usepackage{pst-all}
\usepackage{pstricks-add}
\usepackage{float}

\usepackage{pgfplots}

\usepackage{tikz}
\usepackage{tikzscale}
\usepackage{transparent}
\usetikzlibrary{snakes}
\pgfplotsset{
  /pgfplots/line legend with two nodes/.style 2 args={
    legend image code/.code={
      \draw[##1,no markers]
        plot coordinates {
        (0cm,0cm)
        (0.3cm,0cm)
        (0.6cm,0cm)
      }
      node[pos=0,#1]{}
      node[#2]{};%
    }
  }
}

\usepackage{algorithm}
\usepackage[noend]{algpseudocode}

\usepackage{comment}
\def\BibTeX{{\rm B\kern-.05em{\sc i\kern-.025em b}\kern-.08em
    T\kern-.1667em\lower.7ex\hbox{E}\kern-.125emX}}
    \usepackage{mathtools}
    \DeclarePairedDelimiter{\ceil}{\lceil}{\rceil}
\begin{document}


\title{A Social IoT-driven Pedestrian Routing Approach during Epidemic Time}
\author{
\IEEEauthorblockN{Abdullah Khanfor,
  Hamdi Friji,
  Hakim Ghazzai,
  and Yehia Massoud\\
\IEEEauthorblockA{School of Systems \& Enterprises, Stevens Institute of Technology, Hoboken, NJ, USA} 
}
}

\maketitle

\begin{abstract}
The unprecedented worldwide spread of coronavirus disease has significantly sped up the development of technology-based solutions to prevent, combat, monitor, or predict pandemics and/or its evolution. The omnipresence of smart Internet-of-things (IoT) devices can play a predominant role in designing advanced techniques helping in minimizing the risk of contamination. In this paper, we propose a practical framework that uses the Social IoT (SIoT) concept to improve pedestrians safely navigate through a real-wold map of a smart city. The objective is to mitigate the risks of exposure to the virus in high-dense areas where social distancing might not be well-practiced. The proposed routing approach recommends pedestrians' route in a real-time manner while considering other devices' mobility. First, the IoT devices are clustered into communities according to two SIoT relations that consider the devices' locations and the friendship levels among their owners. Accordingly, the city map roads are assigned weights representing their safety levels. Afterward, a navigation algorithm, namely the Dijkstra algorithm, is applied to recommend the safest route to follow. Simulation results applied on a real-world IoT data set have shown the ability of the proposed approach in achieving trade-offs between both safest and shortest paths according to the pedestrian preference. 
\end{abstract}


\blfootnote{\hrule
\vspace{0.2cm} This paper is accepted for publication in 2020 IEEE Global Conference on AI \& IoT IEEE (GCAIOT'20), Dubai, UAE, Dec. 2020. \newline \textcopyright~2020 IEEE. Personal use of this material is permitted. Permission from IEEE must be obtained for all other uses, in any current or future media, including reprinting/republishing this material for advertising or promotional purposes, creating new collective works, for resale or redistribution to servers or lists, or reuse of any copyrighted component of this work in other works.}%

\section{Introduction}
In early 2020, the world was hit by an unprecedented pandemic that severely affected most countries and created global health care and economic pressures. To prevent its spread, limiting the exposure to the virus is the main priority of local authorities. Precautionary practices such as hand cleaning, mask-wearing, social distancing, and close contact avoidance are highly recommended and even imposed in many countries. Besides, technological solutions have been tested and implemented to help mitigate the spread of COVID-19 in the world. One of the most promising approaches is to exploit heterogeneous and omnipresent communication systems such as the Internet of Things (IoT) to enable e-monitoring techniques such as spread tracking, contact tracing, and crowded areas monitoring. IoT can provide cost-efficient and practical solutions to help practicing social distancing and hence, limit the spread of the infection~\cite{singh2020internet,rahman2020defending}.


The current smartphones and wearable devices' and the existing infrastructure can boost the development of IoT-based solutions in a quick and large-scale manner to combat pandemics. Many examples of IoT-based solutions to contend the pandemic effects have been proposed in literature~\cite{chamola2020comprehensive}\cite{pimpinella2019walk}\cite{hatzopoulou2013web}. For instance, the smart disease surveillance systems had demonstrated an efficient degree of control for the pandemic's spread within the city of Wuhan and other major cities in China~\cite{peeri2020sars}. Despite the significant issues of privacy, South Korea's exemplary accomplishments for containing the COVID-19 until today is due, in part, to the commissioning a coherent information system that tracks visitors and confirmed patients with an alerting system of potential infections~\cite{ting2020digital}. The system provides the community with essential information to assess the spread. Taiwan used various IoT technologies, such as tracking the citizens and travelers through their mobile phone locations. Thus, if citizens are exposed to an area with a high risk of getting infected, they will be altered. Also, if a traveler comes from a high-risk area and violates self-quarantine procedure during the incubation phase, in that case, the residents in that area will be notified through a text message to alert them~\cite{wang2020response}.

A fundamental solution to reduce the transmission of infectious diseases in general and particularly the COVID-19 is maintaining social distancing. IoT can perform a vital function in helping with social distancing practices. Thus, the built-in capabilities of connected devices such as GPS, thermometer, and other sensors in the IoT system can help in social distancing. For instance, in construction or industrial zones, wearables can be employed to maintain a safe distance between workers by generating alerts if social distancing is violated. It also helps track the spread of the virus in case that an infected person was present in the working area and hence, avoid the complete shutdown of the institution~\cite{ting2020digital}. The emergence of social IoT (SIoT) can be a valuable tool to leverage the traditional IoT systems and enable a better understanding of the ubiquitous IoT network~\cite{khanfor2019application}. SIoT model the devices and users in the system with different social relations interconnecting the IoT devices. These relationships can be established between machine-to-machine, human-to-machine, and human-to-human connections~\cite{atzori2012social} and transform the IoT network into a socially connected network of devices that can be effectively analyzed using graph analytics tools such as community detection \cite{shuja2020applying} and machine learning \cite{khanfor2020computational}. By assessing the SIoT, providing new applications to battle the virus spread can emerge and contribute to minimizing the pandemic's negative impacts.




\begin{figure*}[t]
\centerline{\includegraphics[width=0.95\textwidth]{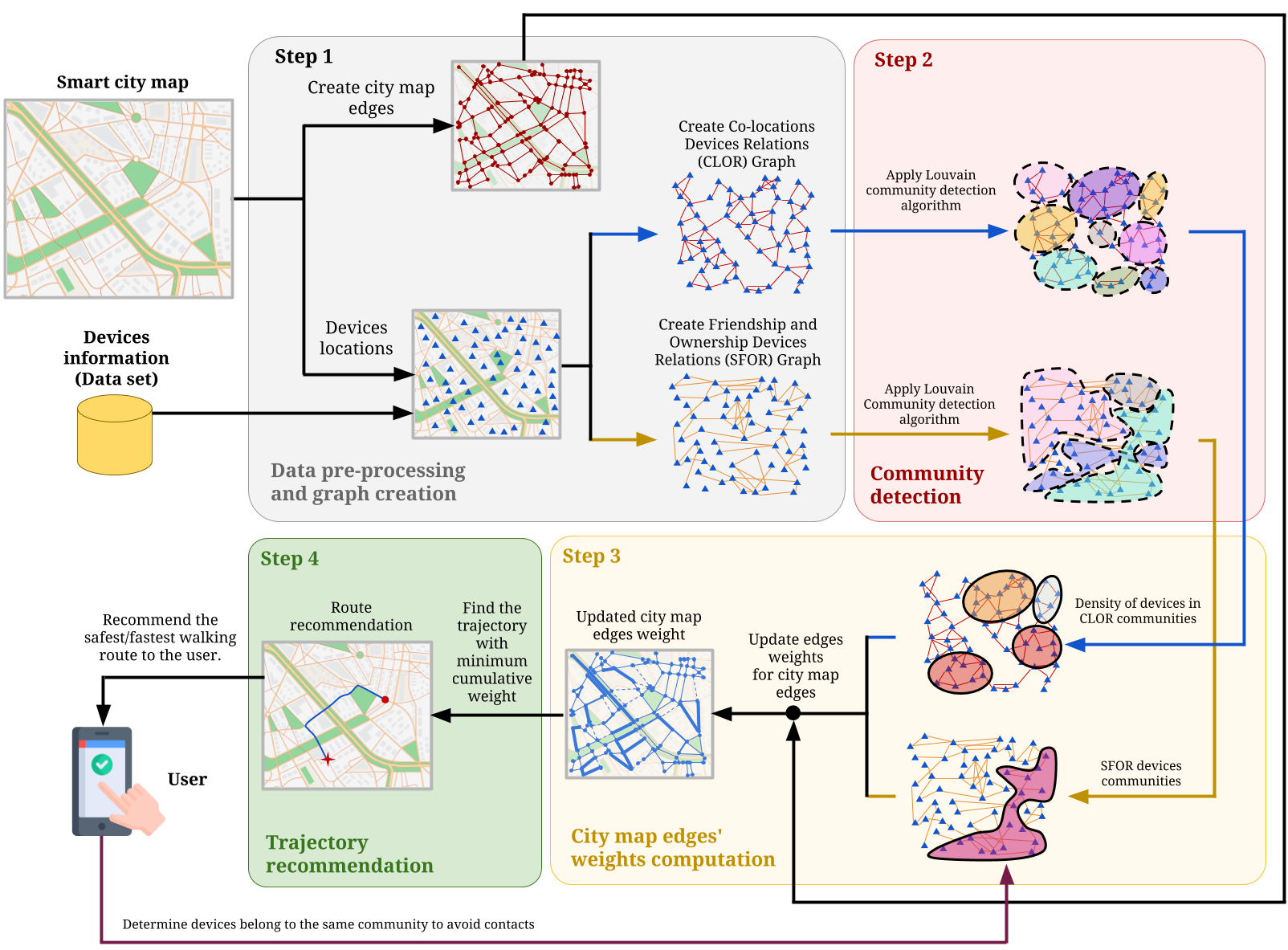}}
\caption{A proposed framework to recommend a safe and fast route to the user during a pandemic in SIoT.}
\label{fig:framework}
\vspace{-0.2cm}
\end{figure*}

In this paper, we propose a smart navigation framework intending to determine for pedestrians safe routing to bypass areas where the risk of COVID-19 transmission is high. In other words, the framework recommends a pedestrian walking route in which guarantees a social distancing and avoiding close contacts. The proposed approach includes four steps: First, the framework identifies the IoT devices located in the area of interest and then establishes social graphs interconnecting these devices using different social IoT relations. Then, the Louvain community detection algorithm is applied to the SIoT graphs to determine different communities of IoT devices. In our approach, we focus primarily on two social relations: a distance-based relation that identifies crowded/high-density areas of IoT devices and a device friendship relation that allows labeling streets where the user may possess a high chance of meeting a close friend. The third step is to compute different scores representing each street's safety level or segment of a street in the area of interest according to the nearby detected social communities. Finally, in the last step, the city map is transformed into a weighted undirected graph to which we apply the Dijkstra algorithm \cite{dijkstra1959note} in order to determine a route characterized by a certain level of safety. A weighted bi-objective function balancing between the shortest and safest routes is developed and implemented. The framework will then deliver the trajectories to the user, e.g., via a mobile application for the best route to follow to reach a destination. The proposed routing approach takes into account the mobility of IoT devices and may update the recommended route regularly by repeating the process, as mentioned earlier.


\section{Proposed Pedestrian Routing Approach}
In Fig.~\ref{fig:framework}, we present a flowchart of the proposed navigation framework where the four steps are showcased. The objective is to recommend a safe route for a user connected to the server through its IoT device, e.g., a smartphone. To this end, two inputs are required to determine the route: the offline city map and a data set containing the IoT device information such as the device locations, and device owners. The framework will output a route from a starting point $A$ to a destination $B$ that is already pre-defined by the pedestrian of interest. The first step is a pre-processing step in which three graphs are generated: i) a graph representing the city road map, ii) a graph representing the social relation reflecting the geographical relation of the connected IoT devices called the Co-LOcation based relation (CLOR), and a graph identifying the friendship levels among the IoT devices, called the Social Friendship and Ownership-based Relation (SFOR). The second step aims to understand IoT devices' social relations better and relax the problem's complexity by determining communities of socially connected IoT devices. To this end, we propose to employ a community detection algorithm, namely the Louvain method~\cite{blondel2008fast}, on the CLOR and SFOR graphs. This step will output communities with different risk levels of virus exposure. In the next step, step 3, we assign to every edge of the city map graph a weight measuring the traveled distance as well as the safety level of the corresponding street or segment of the street. Finally, the last step applies a graph routing algorithm that minimizes the weights along the selected trajectory and determines the best path to recommend to the pedestrian of interest. In the following, we explain in detail each step of the proposed navigation framework.

\subsection{Step 1: Data Pre-processing and Graph Generation}
In order to adequately manipulate the road network of the geographical area of interest, we convert it into a graph where the vertices correspond to the intersections of the roads or connections of two consecutive segments of roads. Indeed, we divide long roads/streets having lengths higher than a threshold length $L_{th}$ into multiple segments to each segment of road is treated solely and might be later assigned different weights. The number of segments for a road of length $L_{road}$ can be expressed as $\ceil{\frac{L_{road}}{L_{th}}}$. Hence, the edges of the graph are the obtained segments of roads connecting two consecutive connections or the ones connected with an intersection.

Besides, we generate two other graphs related to the different social relations and based on the available data set of the connected IoT devices. In the following, we list two different relations that are used in this study to determine to measure the social interconnections among the IoT devices:\\
$\bullet$ \textit{Co-location/co-work based relation (CLOR):} The geographical locations of the IoT devices can be used to define a specific relation reflecting the fact that two devices are co-located in a given area at a certain instant of time. By setting a defined threshold for the distance between the devices, we can specify whether these devices belong to a specific cluster or not and hence, establish relations between them based on their separating distances. This relation will allow identifying crowded areas where there is a risk that social distancing is not practiced among these devices and hence, it is vital to avoid passing by these hot spots.\\
$\bullet$ \textit{Social friendship and ownership relation (SFOR):} In this relation, we identify the devices that might be owned by the pedestrian of interest of his/her social friends. It is assumed that there is more chance that the pedestrian will meet or be in close contact with people using these devices, and hence, there is a risk of contamination with people that he/she knows. To create SFOR relations, we first consider that two devices owned by the same person are strongly connected. Regarding devices owned by different owners, we establish their SFOR relations using social media networks or other friendship indicators. For example, if two owners are friends in the social network, then the SFOR relation between their devices can be modeled by an edge with a weight reflecting the strength of their relations. SFOR relation can be extended to the case of a friend of a friend with a reduced weight computed according to the number of friends needed to reach a specific device.

The CLOR and SFOR topologies are undirected and weighted networks. The nodes in these graphs are heterogeneous IoT devices. The edges between these devices representing the SIoT relations stated previously. These graphs do not include self-loop edges because the relations definitions do not require such a feature.

\subsection{Step 2: Community Detection}
This step focuses better on analyzing the social IoT relations, reducing the complexity of the problem, and serve in information retrieval. A community detection algorithm converts the complex social graphs into clusters of devices sharing strong relations. To this end, we apply the Louvain method~\cite{blondel2008fast}. The main advantage of using the Louvain is the running time of $O(n \log n)$, which is considerably faster comparing to a similar methods~\cite{khanfor2019application,khanfor2020automated}. The outcomes of the community detection in our framework will be used in the next \textit{Step 3} based on the relationships described in \textit{Step 1}, namely CLOR and SFOR.

Applied to the CLOR graph, the Louvain method is expected to extract co-located devices communities. Devices belonging to the same community are considered to be positioned near one another and may create a highly dense zone that is risky to cross through. In SFOR communities, the devices are not necessarily co-located. On the contrary to CLOR, they might be sparsely distributed in a geographical area. However, the owners of these devices may know each other and can meet each other. Therefore, for safety reasons, it is recommended that a given pedestrian do not pass by devices belonging to the same SFOR community of its device.
\begin{figure*}[!t]
    \centering
    \begin{subfigure}{0.43\textwidth}
        \centering
        \fbox{\includegraphics[width=0.9\textwidth]{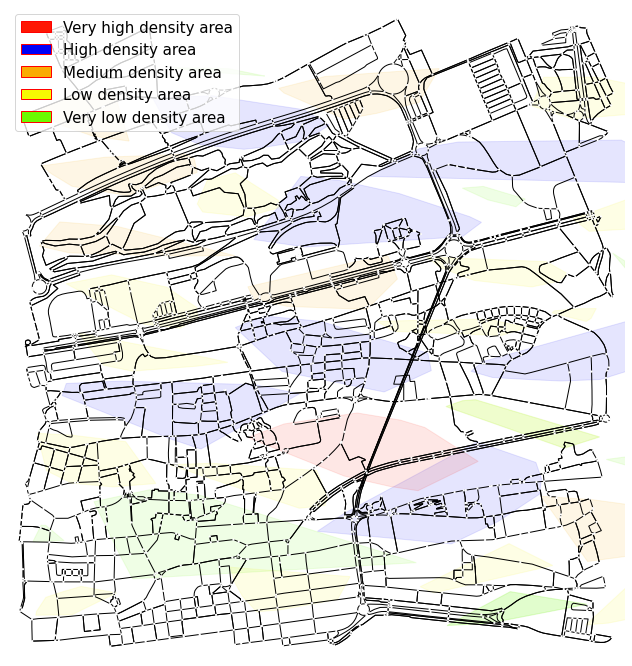}}
        \caption{CLOR communities (The colors of each CLOR community indicates the density level of the community).}
             \label{fig:commCLOR}
    \end{subfigure}
    ~\begin{subfigure}{0.43\textwidth}
        \centering
        \fbox{\includegraphics[width=0.9\textwidth]{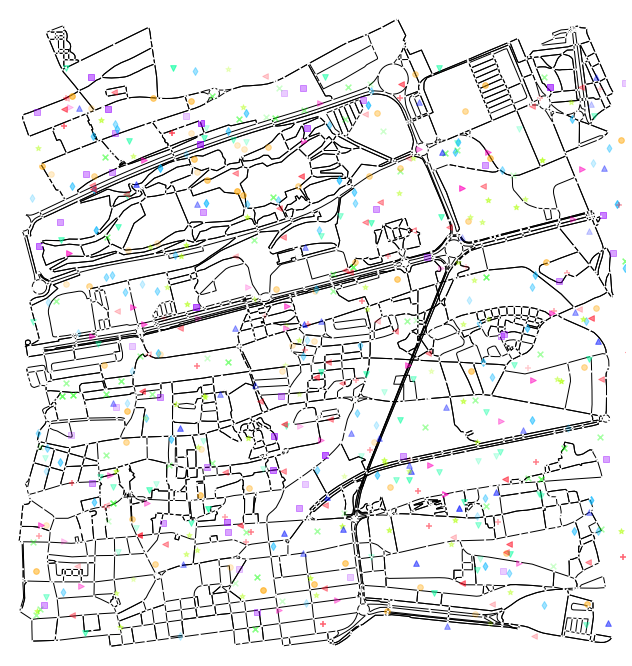}}
        \caption{SFOR communities (IoT devices belonging to the same community are labeled by the same marker).}
        \label{fig:commSFOR}
    \end{subfigure}%
    \caption{Communities were detected based on CLOR and SFOR relations using the Louvain Method.}
\label{fig:communities}
\vspace{-0.5cm}
\end{figure*}

\subsection{Step 3: City Map Edges' Weights Computation}
Since IoT devices are usually omnipresent, it is unlikely to find routes free of devices, i.e., zero-risk zones. Therefore, in this step, we propose to compute weights and assign them to different edges of the road map graph given the statuses of their surrounding communities. Hence, the route selection algorithm will minimize the sum of weights along the route. In this step, we calculate the edges' weights defined as a function balanced by a coefficient $\alpha \left(\in [0,1]\right)$ representing the level of safety set by the pedestrian of interest. Setting a value of $\alpha \rightarrow 0$, the pedestrian intends to determine the shortest path to reach the destination with low consideration of risks. However, if $\alpha \rightarrow 1$, the pedestrian is looking to follow the safest trajectory independently of the expected traveled distance. Values of $\alpha \in ]0,1[$ achieves a trade-off between both routing strategies. The edge's weight of the road network, denoted by $\omega_e$ can be expressed as follows:
\begin{equation}
    \label{equ:weights}
    \omega_e= (1-\alpha)\,\omega_e^{dist} + \alpha\,\omega_e^{sft},
\end{equation}
Where $\omega_e^{dist}$ is the weight reflecting the expected traveled distance when crossing edge $e$, while $\omega_e^{sft}$ is the weight reflecting the safety level of the edge. The weight $\omega_e^{dist}$ is a normalized value of its length. On the other hand, the value of $\omega_e^{sft}$ is obtained by combining the impact of the surrounding CLOR communities and devices belonging to the same SFOR community of the pedestrian's device as follows:
\begin{equation}
    \label{equ:sftweights}
    \omega_e^{sft}=\omega_e^{CLOR} + \omega_e^{SFOR}.
\end{equation}
The CLOR weights $\omega_e^{CLOR}$ are calculated using the following expression:
\begin{equation}
\label{equ:CLOR}
\omega_e^{CLOR} = \sum\limits_{c\,\in\,\mathcal C_e^{CLOR}} \frac{\gamma_c}{|\mathcal C_e^{CLOR}|},
\end{equation}
Where $\mathcal C_e^{CLOR}$ is the set of CLOR communities that intersect with the edge $e$. The cardinality of this set is denoted by $|\mathcal C_e^{CLOR}|$. The CLOR communities are modeled as polygons that circumscribe all devices belonging to them with an outer offset $ \rho $. We denote these polygons by $P_c, \forall c \in \mathcal C^{CLOR}$ where $\mathcal C^{CLOR}$ denotes the set of all CLOR communities obtained using the Louvain method. Hence, the set $\mathcal C_e^{CLOR}$ can be defined as: $\mathcal C_e^{CLOR}=\{ c \in \mathcal C^{CLOR}\,|\, P_c \cap \{e\} \neq \emptyset \}$ where $\emptyset$ is the empty set. Note that the offset parameter $\rho$ is added to all polygons associated with the CLOR communities to ensure a safe social distance separating the navigating user and the devices at the edges. Finally, $\gamma_c$ denotes the density of community $c$ and is calculated as follows:
\begin{equation}
\label{equ:transect}
\gamma_c = \frac{N_{c}}{A_{c}},
\end{equation}
Where $N_c$ is the number of devices in community $c$ and $A_c$ is the surface of the area of $P_c$.

Similarly the SFOR weights can be computed as follows:
\begin{equation}
\label{equ:SFOR}
\omega_e^{SFOR} = \sum\limits_{u \in \mathcal C_{u^*,e}} \frac{\Omega^{SFOR}_{u,u^*}}{|\mathcal C_{u^*,e}|},
\end{equation}
Where $\mathcal C_{u^*,e}$ is the set of devices that are in SFOR relation with the device $u^*$ of the pedestrian of interest and their $d(u,e)\leq d_{th}$. It corresponds to the SFOR community to which the device $u^*$ belongs having a distance $d_{th}$ or less from the edge $e$. The coefficient $\Omega^{SFOR}_{u,u^*}$ is obtained from the SFOR graph, and it measures the SFOR relation between device $u$ and $u^*$. The parameter $d_{th}$ can represent a distance from which the users owning the IoT devices cannot see each other and hence, do not meet and avoid close contact.  

Notice that the safety weight of an edge $e$ significantly increases if it is surrounded by high-density CLOR communities and/or many devices belonging to the same SFOR community of the user of interest. Therefore, in the next step, we aim to select the edges, i.e., the trajectory that minimizes the sum of $w_e^{sft}$ for a user looking for a safe walk.

\subsection{Step 4: Trajectory Recommendation}
After computing the weights of the city map graph in \textit{Step~3}, we apply the Dijkstra's shortest path algorithm to determine the trajectory with minimum cumulative weights of the selected trajectory between the points $A$ and $B$. We propose to employ Dijkstra's algorithm due to its reasonable running. However, similar algorithms can be applied in our context.

Our framework is capable of dynamically recommending new paths to the user based on his/her current location while considering the mobility of other IoT devices. Hence, it needs to update the CLOR and SFOR communities after a specific period. Consequently, the selected path is updated from a time slot to another. In other words, the steps 2 to 4 will be repeated for each time slot until the user reaches his/her destination.

\section{Results \& Discussions}\label{sec:results}
\begin{figure*}[!t]
    \centering
    \centerline{\includegraphics[width=0.95\textwidth]{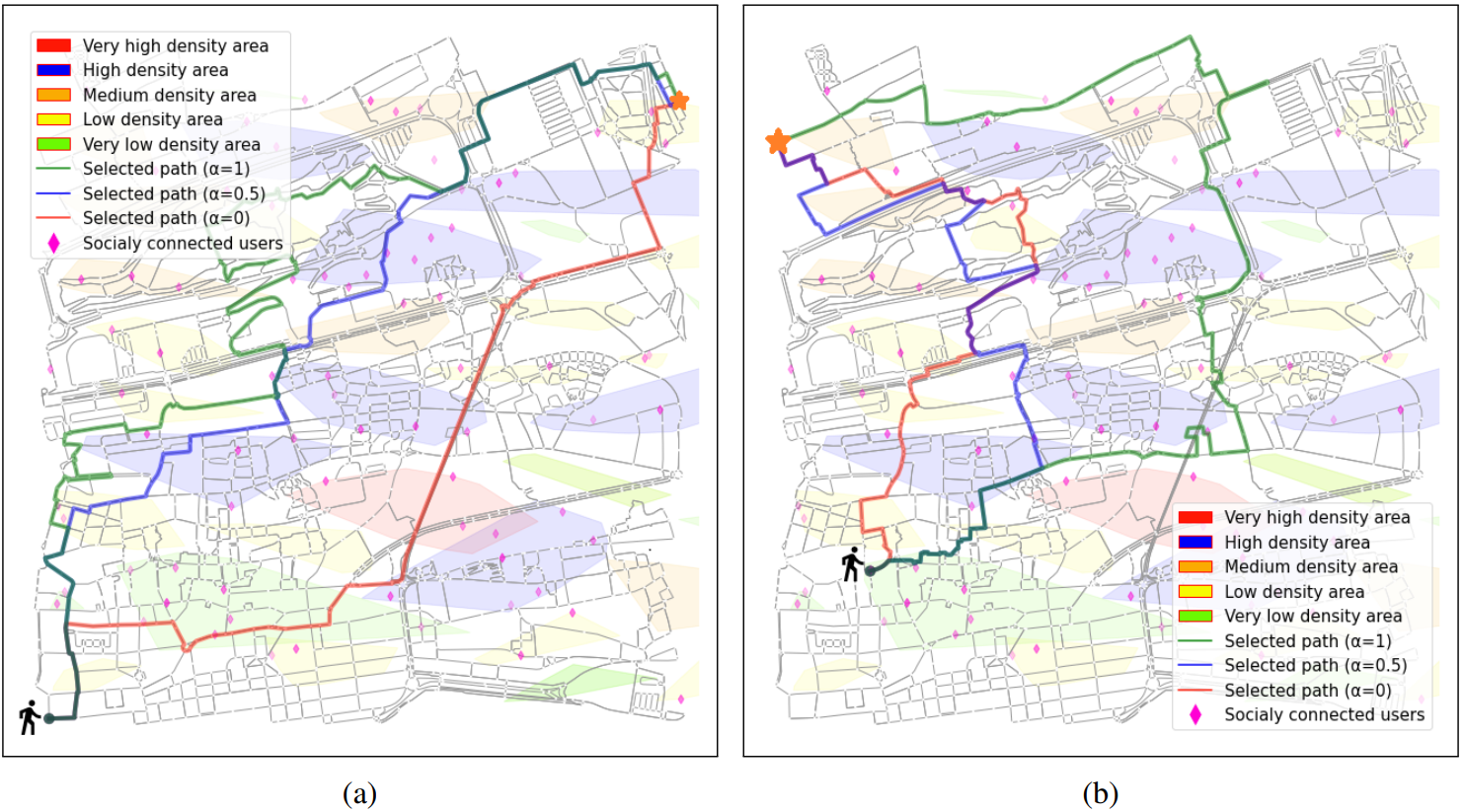}}
    \caption{Two examples showing the different paths recommended to the user for different values of $\alpha$.}
\label{fig:performance}
\vspace{-0.5cm}
\end{figure*}

In our simulations, we select a $6 \times 6$ km$^2$ area in Santander, Spain. We extract the map using OpenStreetMap project\footnote{https://www.openstreetmap.org} and convert it to a road graph using the OSMnx method~\cite{boeing2017osmnx}. Moreover, we project the devices in the selected area from a real-world IoT data set provided in~\cite{marche2018dataset}. The data set includes 16216 devices covering the whole city. The devices vary from simple sensors such as street lights, environment sensors, and highly computational devices such as smartphones and personal computers. The devices are owned by private and public entities. The local authorities usually own public devices. For the private-owned devices, there are static and mobile devices. In our simulations, we select mobile devices owned by private entities that are most likely owned by human beings such as smartphones, smartwatches, tablets, personal computers, etc. The remaining devices from the previous selection process result in 1312 personal IoT devices.

The selected devices will have SFOR and CLOR SIoT relations. For the CLOR relations, we create a mesh network and drop the edges connecting two devices separated by a distance higher than 1 km. The community detection algorithm applied to the CLOR will return a set of relatively high-density communities to determine high-risk infection areas with the limited practice of social distancing. For the SFOR relation, we employ the social network of the owners of the IoT devices. Since we lack access to the owners' social network, we use Watts–Strogatz generator~\cite{watts1998collective} that portrays a social network between the owners. The relations of devices with the same owner are assigned an edge of 1, while the direct friends-owned devices will have an edge of 0.5. Other devices are given weights computed while considering the minimum number of hops needed for one of the vertices (owners) to reach the other vertices. We restricted the relations to three friends of friends since they are socially far away from each other.

\begin{figure*}[!t]
    \centering
    \centerline{\includegraphics[width=0.95\textwidth]{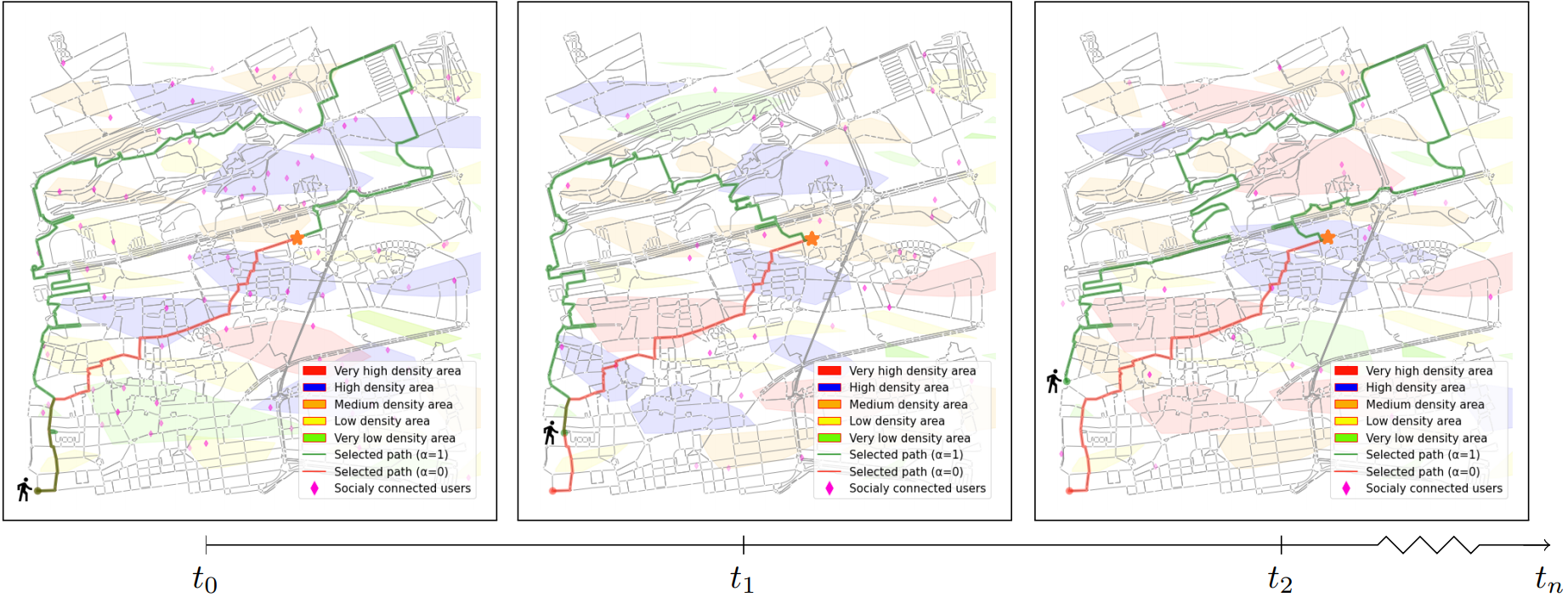}}
    \caption{The routing updates on three timesteps are based on the changes in device locations and communities in the graph.}
\label{fig:dynamic}
\vspace{-0.2cm}
\end{figure*}

Fig.~\ref{fig:communities} shows the communities obtained from the SIoT graphs by applying the Louvain method. We obtain 56 CLOR communities represented by colored polygons and having different density levels, as illustrated in Fig.~\ref{fig:commCLOR}. The CLOR communities are classified, based on their densities, into five classes. There is one very high-density community located almost at the center of the map and other blue CLOR communities with high-density that the user needs to avoid for safety. The medium and low-risk areas might be avoided, but they can be recommended. For the SFOR, it results in 10 communities with diverse types of devices located all over the map. Each community is denoted with different shapes and colors in Fig.~\ref{fig:commSFOR}. The user of interest will belong to one of these communities and needs to avoid close contact with them.

Fig.~\ref{fig:performance} illustrates two examples of the recommended routes for the user given different starting points and destinations for three values of $\alpha$ after applying the Dijkstra's algorithm using the computed weights. If $\alpha=0$, then the framework will recommend the shortest path (the red route), otherwise if $\alpha=1$, the safest path with minimum exposure to the virus is recommended (the green). However, for $\alpha=0.5$, a trade-off between both metrics is provided (the blue route). In Fig.~\ref{fig:performance}, it is clear that along the green route, the user is avoiding most of the high dense areas by surrounding them. It just crosses some of the low-density areas. It also avoids getting closer to other SFOR-related devices unless it is forced to do it. This leads to a long route of 2.77 km. With the red route, the user is unaware of the risk of contamination and crosses all the high-density areas. The corresponding traveled distance is equal to 2.1 km. Finally, for the blue route, the algorithm avoids the red zone and tolerates passing by some blue areas. This results in a traveled distance of 2.4 km.
\begin{figure}[t]
    \centering
    \centerline{\includegraphics[width=\columnwidth]{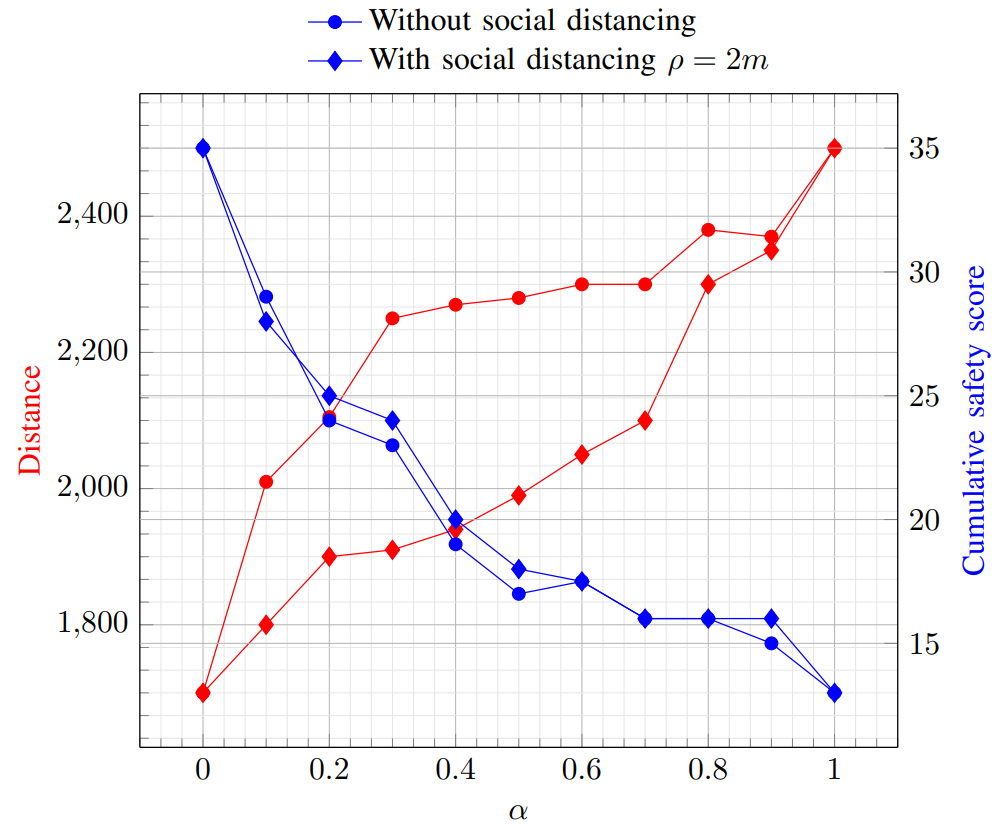}}
    \vspace{-0.2cm}
\caption{A trade-off between the safety factor and destination amount for the proposed framework.}
\label{fig:alphatradeoff}
\vspace{-0.5cm}
\end{figure}

Our framework can be adapted to a dynamic scenario by examining real-time mobility. In Fig.~\ref{fig:dynamic}, we plot the routes for three consecutive time slots. Each time slot, the IoT devices change their locations and, consequently, influence the CLOR relations and their communities and the position of devices in SFOR. In the dynamic scenario, the shortest path will remain intact. The algorithm is only aware of the distance to be crossed, while in the proposed framework that considers the safety weights, the trajectory is regularly updated given the location of the devices at each time slot. Accordingly, the user starting at the left bottom corner of the map will notice that its trajectory is partially updated at a time slot ($t_1$) since several SFOR devices left the area, and the user can cross in the middle of the map to reach its destination. In the next time slot ($t_2$), the navigation algorithm is executed again. Notice that the user is forced to go around it to reach its destination. As long as he/she is moving, the user is getting closer to the destination, especially if a correct value of $\alpha$ is chosen. Choosing $\alpha$ close to 1 may lead to confusing results. Therefore, to balance between safety and travel time, an optimized choice of $\alpha$ should be made.

In Fig.~\ref{fig:alphatradeoff}, we plot the final travel distance and a safety score measuring the cumulative safety weights along the trajectory versus different values of $\alpha$ for two choices of $\rho$ (the social distancing outer offset). A higher value of $\rho$ indicates an increasing preventive navigation strategy aiming to find trajectories a little bit far away from CLOR communities. The figure shows that by increasing $ \alpha $, the travel distance increases, moving from 1.7 km to around 2.5 km while the cumulative safety score is almost linearly decreasing. A compromise between safety and speed can be achieved for $\alpha$ around 0.4. By increasing the outer offset $ \rho $, a more strict social distancing is applied, and hence, the traveled distance increases even for the same value of $ \alpha $. For instance, for $\alpha=0.4$, the distance changes from 1.95 km to 2.21 km, with a slight improvement of the safety score.


\vspace{-0.3cm}
\section{Conclusion}
Ubiquitous IoT can provide solutions to combat pandemics such as COVID-19. We provide a practical framework that could assist in such circumstances. The framework recommends a trajectory for a pedestrian user to reach his/her destination while avoiding areas with high risk of exposure to the virus. It employs technical and social advantages of smart IoT devices to determine risky areas and help people better practice social distancing. The framework can be update the recommended route in real-time based on the user's needs and mobility of other devices. As future work, the framework will be extended to consider the case of multi-user routing and can be adapted to more complicated areas such as indoor and industrial workplaces.

\bibliography{references}
\bibliographystyle{ieeetr}
\balance

\end{document}